\documentclass{article}

\usepackage{makecell}
\usepackage{upgreek}
\usepackage{amsmath}
\usepackage{graphicx}
\usepackage{url}

\newcommand{\mce}[1]{\hskip-1.3mm\makecell{#1}}      

\newcommand{\mces}[1]{\small{\hskip-1.3mm\makecell{#1}}}      

\newcommand{\eexp}[1]{\mbox{$\hskip0.9pt\times\hskip0.4pt10^{#1}\hskip2.2pt$}}
\newcommand{\eeexp}[1]{\mbox{$\hskip0.7pt\cdot10^{\hskip0.4pt #1}\hskip2.2pt$}}

\newcommand{\captionss}[1]{\caption{\small{#1}}}

\newcommand{\sss}[1]{\mbox{\hskip1.9pt\ref{#1}}}
\newcommand{\rrr}[1]{\mbox{Figure\hskip1.6pt\ref{#1}}}
  \newcommand{\ttt}[1]{\mbox{Table\hskip1.9pt\ref{#1}}}

      \newcommand{\roz}[1]{\mbox{Section\hskip1.3pt\ref{#1}}}

\newcommand{\tsea}{\hspace{1mm}}
\newcommand{\tseb}{\hspace{2mm}}   
\newcommand{\tsec}{\hspace{3.5mm}}

\newcommand{\pusline}[1]{\rule[0mm]{0mm}{#1}} 

\begin{document}

\title{Venus and Bright Planets Through a Large-Scale Camera Obscura: Optical and Mechanical Issues of Conducting the Observations}

\author{Krzysztof W\'{o}jcik }

\date{
Faculty of Mechanical Engineering, 
Cracow University of Technology \\ Al. Jana Pawla II 37 \\ Cracow, Poland; krzysztof.wojcik@pk.edu.pl \\ \today}

\maketitle


\begin{abstract}
The article describes observations of the crescent of Venus and other bright planets using a large camera obscura.
The goal of the article is to demonstrate that these observations can be successfully performed.
Achieving positive results depends on solving several key problems.
The most significant of these is the difficulty of perceiving the extremely faint light of a planet on the camera obscura screen. This issue is resolved through the use of special directional screens. Two main types (translucent and reflective) are described in the article.
The construction of a so-called ``artificial Venus'', designed to test directional screens and determine the average sensitivity of human vision, is also presented. \\
Other serious challenges include aiming the camera obscura at the celestial object and compensating for Earth's rotation. One of the methods discussed involves the use of a flat intermediate mirror and a special mount for its guidance.\\
The objective of the paper is fulfilled through the presentation of both visual and photographic observation results. In addition to Venus's crescent, observations of the Saturn's rings are also presented.
The proposed design of the camera obscura, which incorporates a specialized projection system, enables the separation of its two functions: directing light from  objects and focusing its energy.
The elements performing these tasks are fully scalable. This can be used in the construction of modern telescopes.
The article also comments on the possibility of observing the phases of Venus in the distant past and the important consequences resulting from this.
\end{abstract}


\section{Introduction}             
 \label{intro}  

Venus draws attention with its rapid movement across the celestial sphere and its intense brightness.
The planet intrigued our ancestors as well, as evidenced by its presence in world religions, cultures, and philosophies. 
Since ancient times, people have tried to fit Venus's movements into models of the universe they have built.
The question of the origin of its light was considered, and attempts were made to observe its possible crescent \cite{ger2}.
The maximum angular size of the crescent is on the order of 
1' (approximately 0.3\eexp{-3}rad),     
which means that it is at the limit of the resolving power of the human eye.
Under extremely favorable conditions, the observation of the planet's elongated shape is potentially possible.

Trustworthy descriptions of such observations made in ancient times have not survived to the present day \cite{ger1,ger2}.
Conducting a whole series of observations of Venus's phase would indicate that the planet is a sphere illuminated by the Sun \cite{vis,ger2}.
A correct interpretation of these observations would allow for the formulation and justification of the heliocentric model of the universe with its religious and scientific consequences long before Copernicus and the telescopic observations made by Galileo \cite{hitel,ziol}.

The paper has two main goals. The first is to experimentally demonstrate that it is possible to observe the crescent shape of Venus using a camera obscura. The second is to describe the design details of a large-scale camera that enable such observations.
However, we do not attempt to answer the question of whether such observations actually occurred in the past.
Answering this question requires undertaking dedicated historical and archaeological research.

In addition to visual observations made with the camera obscura, we also describe their modern extension in the form of photographic observations. The resulting images illustrate the appearance of Venus's crescent as seen visually.
The satisfactory results of the conducted research prompted the idea of using the camera obscura to observe other bright planets in the Solar System.
These observations show the capabilities of the camera as an astronomical instrument.

We will present the problems of camera obscura construction through a description of four key issues: resolution, the brightness of the created image, aiming the camera at the planet, and compensating for the Earth's rotation. 
These topics are explored in Sections \ref{resol}--\ref{comp}. The issue of using the camera obscura for photographic observations is described in Section \ref{foto}.
Section \ref{obser} presents the results of both visual and photographic observations.
The findings are discussed in Section \ref{discussion}.
Main outcomes of the study are summarized in Section  \ref{concl}.

\section{Resolution of the camera obscura}
\label{resol}
\nopagebreak
Figure\sss{cam} depicts the geometric model of the camera obscura \cite{inside}.
The points of the observed object are projected onto the camera screen in the form of discs.
The symbol $D$ denotes the aperture diameter of the camera, and $f$ denotes the focal length (we will use this term even though the light is not focused in the traditional sense). It is assumed that the distance to the observed object, denoted by $l$, is many times greater than  $f$. 
\begin{figure*}[htbp]           
\begin{center}
\includegraphics[width=0.95\textwidth]{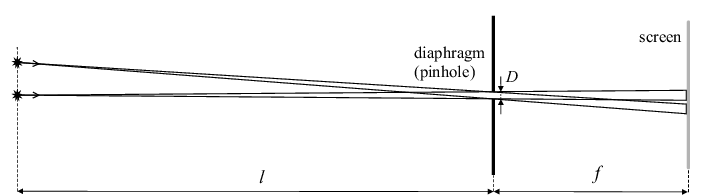}  
\captionss{Image of two points in the camera obscura.}
\label{cam} 
\end{center}
\end{figure*}    

Based on the presented model, the geometric (or aperture) resolution of the camera  can be defined as the angle: 
\begin{equation}
R_{ap} = \beta \frac{D}{f}
\label{rap}
\end{equation}
where $\beta$  is an arbitrarily chosen constant. For example, if we assume $\beta =0.5$, we can interpret $R_{ap}$ as the angular distance between two spaced, luminous points that create images on the screen in the form of overlapping discs. The distance between their centers equals $D/2$.

Reducing the aperture diameter $D$ leads to an improvement in the geometric resolution. Unfortunately, such reduction renders diffraction phenomena more and more visible. The characteristics of the diffraction pattern, which appears on the screen, are used to define resolution measures. We define the diffraction resolution as the angle: 
\begin{equation}
R_{dif} = \gamma \frac{\lambda}{D}
\label{rdiff}
\end{equation}
where $\lambda$ is the wavelength of light and $\gamma$ is a constant (for the value $\gamma = 1.22$, we obtain the Rayleigh resolution criterion, and with $\gamma = 1$, the Dawes criterion \cite{rayl,roy}).

The total resolution, depending on $R_{ap}$ and $R_{dif}$, can be defined in many ways, for example, using a certain metric
$\mathcal{M} \big (R_{ap}, R_{dif} \big) $.          
In the following discussion, we will use the simple Euclidean metric\footnote{In the case of a large relative difference between $R_{ap}$ and $R_{dif}$, the smaller component has little impact on the output metric value. This effect can be adjusted by using the Minkowski metric: 
$ (  R_{ap}^\eta  + R_{dif}^\eta  )^{1/\eta}$; $\, \eta$ -- selected constant. Additionally, appropriate normalization of the components may be applied.}. 
Thus, the total resolution is defined by the expression:
\begin{equation}
R_{t} = \left(  R_{ap}^2  + R_{dif}^2   \right)^{1/2}
\label{rt}
\end{equation}
The resolution $ R_{t}$   is a function of the aperture diameter $D$. 
In our case, this function reaches its minimum value when the $R_{ap}$ and $R_{dif}$ components are equal:
\begin{equation}
\beta \frac{D}{f} = \gamma \frac{\lambda}{D}
\label{rownr}
\end{equation}
This dependence enables the calculation of the optimal diameter of the camera obscura aperture for a given focal length. The equation (\ref{rownr}) can be rewritten using only one parameter. 
Introducing $\alpha = \gamma / \beta$,  we obtain:
\begin{equation}
\alpha =  \frac{D^2}{\lambda f}
\label{ralfa}
\end{equation}
The parameter $\alpha$ determines the relationship between the two resolution components in a camera with focal length $f$ and aperture diameter $D$. Thus, when two different cameras (with different $D$ and $f$) have the same $\alpha$ value, they produce images that are similar to each other (differing only in scale).
Figure\sss{kraz} depicts experimentally obtained images of a point source of white light.
\begin{figure*}[htb]
\begin{center}
\includegraphics[width=0.95\textwidth]{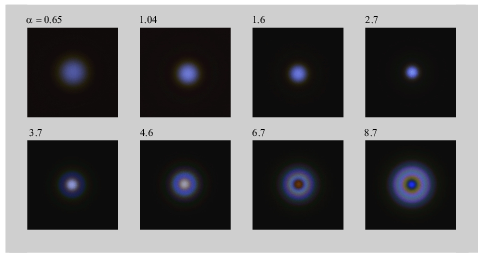}  
\captionss{Images of a point light source on the camera obscura screen obtained for 
different values of the parameter $\alpha$.}
\label{kraz} 
\end{center}
\end{figure*}
The pictures were produced for different values of the parameter $\alpha$ 
(each of the presented images can also be generated numerically using the associated Point Spread Function, PSF \cite{psf2,model}).

Knowing that $\alpha$ determines the type of artifacts produced in the image, we can select its value according to the character of the observed object.
 Most often, we use a value in the range $\langle 2,5 \rangle$. For $\alpha =3.61$, equation (\ref{ralfa}) is equivalent to the Petzval criterion, which determines the optimal aperture diameter for the camera obscura. 
We can note that when $\alpha$ approaches 6, a region of reduced brightness appears in the center (\rrr{kraz}).
This is caused by destructive interference between waves diffracted at the edge of the aperture and those passing directly through it\footnote{Using a simple drawing, this phenomenon can be easily explained theoretically.}. 
In such cases, the PSF can emphasize certain features, such as the narrow crescent shape of Venus.

To conclude, a very important remark: with fixed values of $\beta$  and $\gamma$, and for a camera with an optimal aperture selected so that  $R_{ap} = R_{dif}$, the total resolution $R_{t}$  is inversely proportional to the square root of the focal length:
\begin{equation}
R_{t} = \left(  R_{ap}^2  + R_{dif}^2   \right)^{1/2} = \left(  2 \frac{\beta \gamma \lambda}  {f}\right)^{1/2}
\label{ognis}
\end{equation}
This means that it is theoretically possible to design a camera obscura with arbitrarily high resolution! (we encourage the reader to estimate the focal length required to exceed the resolving power of the best modern telescopes).

Let us get back to the problems with Venus observation.
The angular size of Venus in the near-Earth position is about 
0.3\eexp{-3}rad. Suppose we settle for a resolution of  $10^{-4}$ rad. 
Taking the values: $\beta=0.6, \gamma=1.8$ (i.e., $\alpha=3$), and $\lambda=0.55$\eexp{-6}m,
we obtain: $f=118$~m and $D=13.7$~mm.

\section{Brightness of the image of Venus obtained by the camera}
\label{bright}
\nopagebreak
After a long period of adapting our eyes to seeing in the dark, away from artificial light sources, on a moonless winter night, we can see faint shadows of landscape elements (e.g., trees) created by Venus on the snow \cite{shadows,shad0,urania}. 
Venus's brightness under conditions suitable for observing its crescent is approximately -4.5 mag, corresponding to an illuminance on Earth's surface of
 1.3\eexp{-4}lux.
This value is comparable to the minimum light intensity at which we can observe the contours of objects. Unluckily, the Venus light is further reduced by the camera's aperture. 

Let us treat the camera obscura as a regular photo camera.
The brightness of objects with specific angular sizes on the camera screen is directly proportional to the square of the relative aperture, defined as $D/f$.
However, when observing Venus through a camera obscura, we encounter a situation in which the observed object (Venus) has the angular size comparable to the angular size of the aperture seen from the selected location on the screen.
The given relationship between brightness and aperture is therefore approximate.

Let us consider an ``observer'' located on the screen surface of the camera obscura, within the image of the planet. 
Looking toward the aperture, the observer receives only a portion of the luminous flux that would otherwise pass through without the aperture. This is illustrated in \rrr{przez}. 
\begin{figure}[htb]         
\begin{center}
\includegraphics[width=0.5\textwidth]{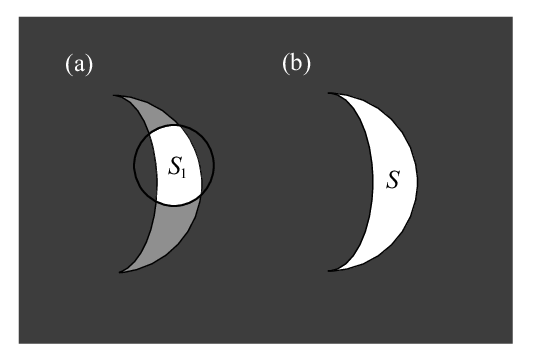}    
\captionss{The aperture-caused reduction of the camera obscura screen illuminance: a) the light beam is limited by the camera aperture, b) without aperture.}
\label{przez} 
\end{center}
\end{figure}
Neglecting diffraction effects, this portion can be approximated by the ratio of the areas: $S_1/S$ (\rrr{przez}).
In our example, depending on the position of the observer within the image of the planet, the ratio described above varies from approximately 1/4 to 1/10 (at the edges of the crescent image). The given values determine the degree of reduction in light intensity on the screen.
As a result of this reduction, observing the Venus crescent becomes nearly impossible.

We need a radical solution to increase the amount of light reaching the observer's eyes. The solution is to use a directional screen. To explain its work, we will compare it with a normal screen made, for example, of a sheet of white paper.
\begin{figure*}[htb]
\begin{center}
\includegraphics[width=1.0\textwidth]{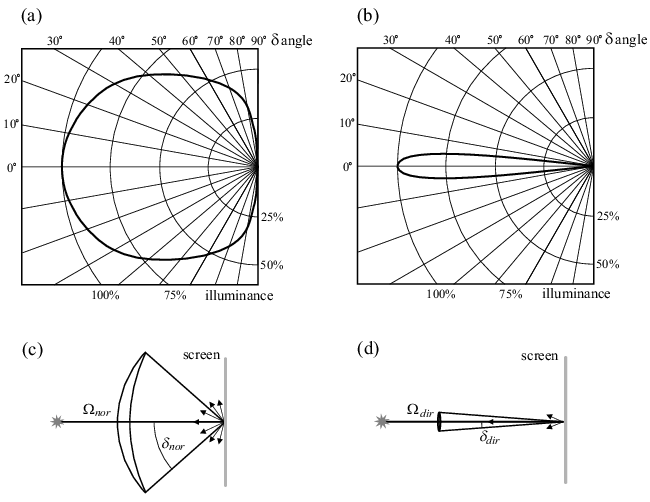}   
\captionss{Comparison of normal and directional screens: (a) light intensity as a function of the angle of reflection -- normal screen, (b) light intensity -- directional screen,  (c) solid angle for the normal screen, (d) solid angle for the directional screen.}
\label{rozprosz} 
\end{center}
\end{figure*}

Suppose both screens absorb about 20\% of the incoming light flux. Further, let us assume that the intensity of light reflected by a normal screen varies with on the angle of reflection $\delta$, as shown in \rrr{rozprosz}(a).  
An analysis of the graph shows that about half of the reflected light flux is contained within a cone with an opening angle of $2 \delta_{nor}$, where $\delta_{nor}$  is about $48^\circ$   (\rrr{rozprosz}(c)). 
This means that half of the reflected beam is contained in a solid angle $\Omega_{nor} = 2.08$~sr  (steradians).      
Suppose that for the directional screen (\rrr{rozprosz}(b)), the angle $\delta_{dir}$, corresponding to half of the reflected beam, equals approximately  $4.5^\circ$  (\rrr{rozprosz}.d). This value refers to a solid angle $\Omega_{dir} = 0.019$ sr.         

As is visible, for a directional screen, half of the light flux is distributed over a solid angle that is about 100 times smaller than that of a standard screen.
Thus, when an observer looking at the directional screen is within the cone associated with  $\Omega_{dir}$, he perceives the brightness of the light reflected from the screen as on average 100 times greater than the brightness of the light reflected by a normal screen.
The value of the angle $\delta_{dir}$ can be treated as a measure characterizing the directional properties of the screen.

Directional screens can be manufactured in various ways (many of which were already available in the Middle Ages). For example, the screen can be a metal mirror with embossed small spherical indentations that change the direction of the reflected rays - \rrr{trans}(a). 

Another possible solution is a translucent screen, where the image is viewed from the reverse side -- \rrr{trans}(b).
\begin{figure*}[htb]
\begin{center}
\includegraphics[width=1.0\textwidth]{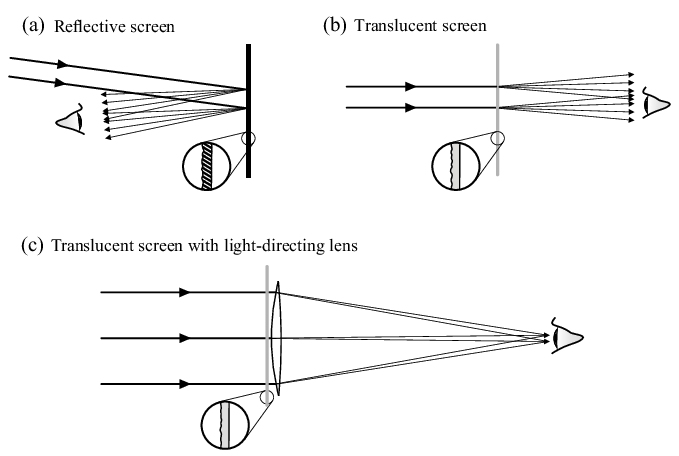}     
\captionss{Types of directional screens.}
\label{trans} 
\end{center}
\end{figure*}
Translucent screens with a small beam spread angle can be constructed by utilizing light refraction on an appropriately corrugated surface.
Such a surface can be created by coating a sheet of glass with a transparent substance (e.g., glue). A transparent PVC film with at least one slightly wrinkled side can also be used.

For screens with a small $\delta_{dir}$, it is possible that rays originating from the edges of the image may not reach the observer's eye.
To address this issue, a lens can be used to direct the light (the observer's eye should be positioned exactly at the focal point of the lens)\footnote{Such light-directing elements are used in telescope and microscope eyepieces.}.
This is illustrated in  \rrr{trans}(c).

A translucent screen with directional properties and slight light scattering can also be achieved using an imprecisely polished lens.
Imperfections on its surface can scatter light at a relatively small angle 
 (e.g.,  $1^\circ$).
In such a  case, the entire light beam may fall into the pupil of the eye (approximately 7 mm in diameter), maximizing the sensitivity of the whole ``screen-eye'' system.

It remains a matter of debate whether such lenses could have been produced in ancient times. 
Currently, of course, we can use ordinary glass or plastic lenses (without corrugated surfaces), as well as Fresnel lenses.

Our considerations regarding the use of directional screens are approximate and require experimental verification.
To this end, a special projection system, the ``artificial Venus'', will be used. It is presented in \rrr{arti}. 
%
%
 
%

\begin{figure*}[htb]
\begin{center}
\includegraphics[width=1.0\textwidth]{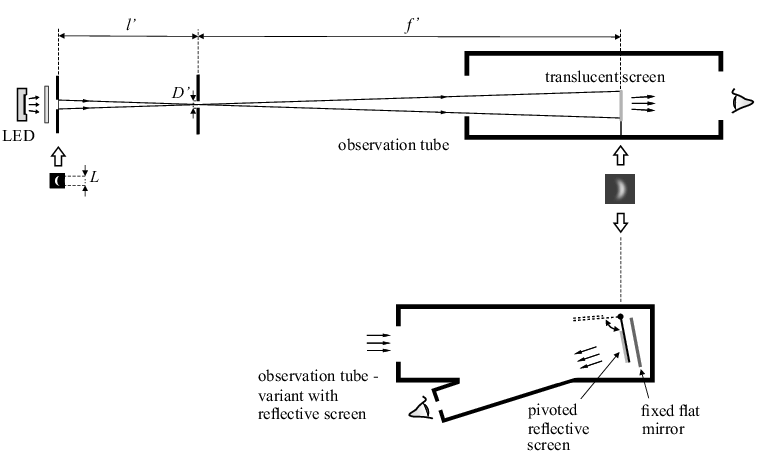}    
\captionss{Diagram of the simulation device (artificial Venus), $l' = 477$~mm, $f' = 5.8$~m, $D' = 0.83$~mm, $L = 1.2$~mm. The equivalent of the $\alpha$ parameter is calculated  according to the equation: $\alpha_{eq}=  D'^2 / \lambda l'$ (this formula is used to ensure that the diffraction pattern produced by the device matches that of the real camera obscura). A version of the observation tube equipped with a reflective screen contains an additional flat mirror. It facilitates the aiming of the tube at the observed object \cite{rep}.}
\label{arti} 
\end{center}
\end{figure*}
The operation of the projector is similar to that of a camera obscura.
It can produce an image on the translucent or reflective screen. The size and shape, including the diffraction pattern, of the produced image correspond to the real image of Venus created by the camera obscura, with parameters $D = 10.5$~mm and $f= 76.2$~m (parameter $\alpha = 2.73$). 
We assumed that the phase angle of the planet is equal to $120^\circ$ (it is defined as the angle  Sun-object-observer). 
The related crescent shape is created by a diaphragm made using classical photographic techniques. The image of the diaphragm is magnified around 12 times.

The electronic components of the device allow the LED brightness to be adjusted. 
The calibration of the LED intensity was made using a digital camera, whose sensor was located in the place of the screen.
The obtained images were compared with actual images of Venus taken using the camera obscura and the same digital camera.

The main purpose of the described projection system is to enable observers to practice the observation techniques related to camera obscura (tube positioning, etc.). This system can also be very easily used to estimate the sensitivity of human vision.

Seven observers (19--56 years old) were asked to determine the direction of inclination of the projected Venus crescent (this proves that the crescent is visible) at various image brightness levels. The translucent screen ($\delta_{dir} \approx 1.7^\circ$) was used.

As a result of the trials, the illuminance thresholds at which the observers correctly indicated the crescent inclination were determined. The results are presented in \ttt{isensi}.
These thresholds are expressed by dimensionless values, whereby the actual illuminance of Venus as seen in the camera obscura refers to 1. 
According to these data, the average illuminance threshold and standard deviation were calculated. They equal 0.162 and 0.136, respectively.

Looking at these results, one can quickly come to the conclusion that human vision has a considerable margin of sensitivity.
Furthermore, using the above data, it is possible to conduct a statistical test that will more reliably confirm that the average threshold defining the sensitivity of our vision is indeed lower than the value corresponding to the real image of the planet Venus.

A Student's t-test was performed, and the obtained value of the Student's t distribution equals $-15.1$. The corresponding p-value equals $1-2.7$\eexp{-6}. 
It should be noted that the condition for applying the Student's t-test is that the analyzed data have a normal distribution.
This requirement was successfully verified using the Shapiro--Wilk test.
Thus, with a very low probability of error (2.7\eexp{-6}), we can confirm the hypothesis that the average sensitivity threshold of human vision is below 
the value corresponding to the actual illuminance of Venus as observed through the camera obscura.

A similarly positive test result was obtained after logarithmizing all analyzed brightness values. This operation corresponds to the work of the sense of sight, 
where the intensity of perception is approximately proportional to the logarithm of brightness (the Weber--Fechner law). 
\ttt{stat} presents the results of the described analyses.

\begin{table}         
\begin{center} 
\caption{ \small{ 
Relative visual sensitivity thresholds for 7 observers (a value of 1 corresponds to the brightness of the actual Venus image) 
}}   
\def\arraystretch{1.1}
\begin{tabular}  {  l  c  c  c  c  c  c  c } 
 No. & 1 & 2 & 3 & 4 & 5 & 6 & 7 \\ 
\hline
Sens. & 0.063 & 0.094 & 0.203 & 0.094 &  0.043 & 0.613 & 0.203 \\ 
\end{tabular}
\label{isensi}  
\end{center}  
\end{table}

\begin{table}         
\begin{center} 
\caption{ \small{Results of statistical analysis of observers' visual sensitivity (critical value for Shapiro--Wilk distribution is 0.803) 
}}   
\def\arraystretch{1.1}
\begin{tabular}  { c  c  c  c }
  & \mce{Stud. t\\distrib.} & \mce{ Stud. t \\ p-value } &  \mce{Shapiro--Wilk \\  distrib. }  \\
\hline
 \mce{ Relative \\value} & $-15.1$ & 2.7\eexp{-6} & 0.820  \\ 
 \pusline{8.0mm}
 \mce{ Logarithmized \\relative \\value} & $-7.26$ & 0.76\eexp{-3} & 0.956 \\ 
\end{tabular}
\label{stat}  
\end{center}  
\end{table}

\section{Pointing the camera at the observed object}
\label{point}
\nopagebreak
A camera obscura with a large focal length can be constructed using natural landscape features such as depressions, caves, valleys, hills, etc. An exemplary design is illustrated in \rrr{teren}(a).

\begin{figure*}[htb]
\begin{center}
\includegraphics[width=0.90\textwidth]{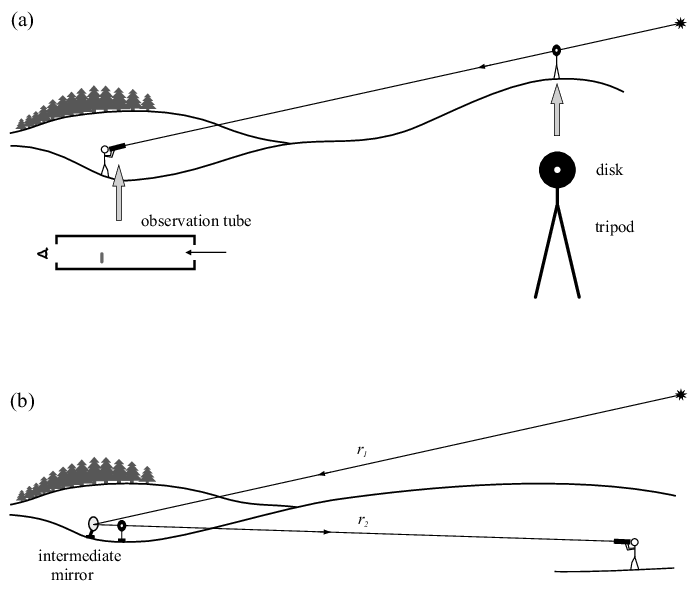}   
\captionss{Construction of a camera obscura using landscape elements:(a) direct observations,(b) observations with an intermediate mirror.} 
\label{teren} 
\end{center}
\end{figure*}

The aperture hole of the camera is created in a special disk, approximately 12 cm in diameter, which also acts as a targeting device. A screen is installed inside an observation tube, constructed analogously to the tubes used in the artificial Venus (\rrr{arti}). 
The purpose of the tube is to suppress scattered light from the night sky.

With the long focal length and the resulting very low brightness of the Venus image, a significant problem is finding it on the screen.
This operation can be carried out in several steps:
The observer, looking through the tube toward Venus, initially sees the planet near the aperture disk without using the translucent screen (the screen only occupies part of the tube's cross-section -- \rrr{arti}). Then, the observer moves together with the tube so that Venus overlaps with the disk and appears in its aperture. 
Finally, we move the tube with the translucent screen in such a way that the Venus light beam passes through it.
A similar procedure applies when using a reflective screen.

The disk, at a great distance from the observer, is poorly visible. To facilitate the targeting process, a small finder scope  mounted on the tube can be used.
Another helpful solution is to install luminous markers (such as LEDs) on the disk. Still another solution is to use several disks or a large array with multiple holes (see supplementary materials \cite{rep}).

During the observation, we must ensure the best possible adaptation of the eyes to the dark. To isolate oneself from any peripheral light, one can use goggles with a small observation hole, special face shields, etc. \cite{rep}. 
If ambient light levels are too high, observations can be performed from inside a special tent, observation dome, or a suitably located building.

\section{Compensating for the Earth's rotation}
\label{comp}
\nopagebreak
%
Another major problem is the movement of the planet's image as a result of the Earth's rotation.
When the image is clearly visible on the screen, the correction motion can be done by the observer himself (he moves along with the observation tube).
In difficult observation conditions, especially when the image is at the limit of visibility, the tube or disk should be moved mechanically, e.g., using devices similar to a slider.

An effective way to compensate for the Earth's rotation is to use a flat intermediate mirror\footnote{A similar technique is used in solar telescopes \cite{roy}.} -- \rrr{teren}(b). 
The mirror may be equipped with a guiding system analogous to that used in telescopes -- \rrr{lustro}.
\begin{figure*}[htb]       
\begin{center}
\includegraphics[width=0.90\textwidth]{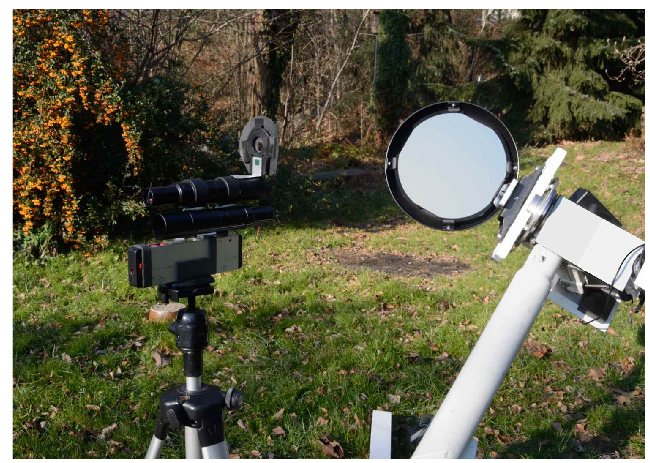}      
\captionss{The mount of the mirror and the camera obscura disk (left of the mirror).
The disk is mounted on an additional instrument to facilitate the initial targeting. The instrument consists of a finder scope  (through which we can observe the planet) and a laser that indicates the location of the observation tube (the laser beam is parallel to the optical axis of the scope). The location of the tube can also be indicated by the second finder \cite{rep}.
} 
\label{lustro} 
\end{center}
\end{figure*}
%
%
We should note, however, that generally, even for equatorial mounts, the orientation of the mirror in space is a non-linear function of time. 
 
In this situation, a feedback loop-based guiding system can be applied \cite{hab}.
One possible solution could be to use an image analysis system.
This system would calculate the position of a bright star or a specific group of such stars seen by the guiding scope and then send appropriate correction signals to the motors that move the mirror (one of the first guiding systems that uses image processing and machine learning methods was presented in detail in \cite{suhora}).

In the observations described in this article, a simple guiding system based on an equatorial mount structure was used. 
The required, typically small, corrections in the declination axis are made by slightly tilting the polar axis east or west. Movement is therefore carried out in a single axis (a single-channel system).
Figure\sss{mech} illustrates the main mechanical components of the mirror mount (similar devices are used in equatorial platforms).
%
%
\begin{figure*}[htb]
\begin{center}
\includegraphics[width=0.97\textwidth]{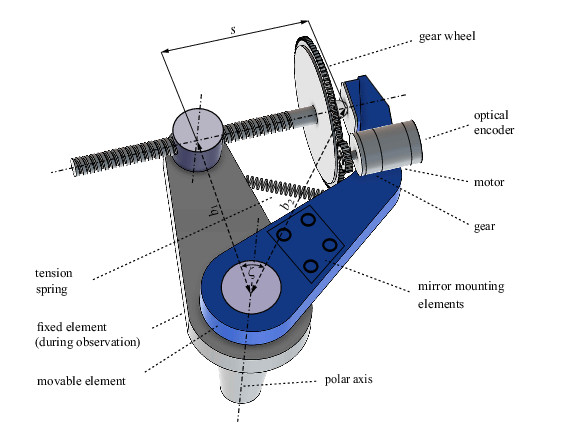}    
\captionss{Mechanical components of the mount system that ensure rotation along the polar axis.
The mirror, with an additional element that enables declination adjustment,
 is attached in the place marked in the drawing.} 
\label{mech} 
\end{center}
\end{figure*}
The guiding system utilizes a DC motor with gear, an optical incremental encoder, and a microcomputer controller\footnote{The system uses a low-power Atmega 328 microcontroller that is based on the AVR RISC architecture.}. These elements work in a local feedback loop.
The system is a discrete nonlinear PID (proportional-integral-derivative) control system \cite{control}.
This design ensures stability\footnote{As defined in control theory \cite{control}.},
 low tracking error, and a relatively short response time. 
Depending on the set angular speed of the mirror, 
the sampling frequency ranges from 20 to 110 Hz.

The system calculates the required rotational speed of the screw with gear wheel  (\rrr{mech}) at successive moments in time. 
Assuming that the mirror orientation in RA (Right Ascension) is a time-dependent function $\zeta(t)$, the distance $s$ (\rrr{mech}) must change non-linearly over time. This change can be determined based on a simple relationship:   
\begin{equation}
s(t) = \left(  b_1^2  + b_2^2 - 2 b_1 b_2  cos^2 \zeta(t)  \right)^{1/2}
\label{nielin}
\end{equation}
where $s(t)$ represents the dependence of $s$ value on time, denotations as shown in \rrr{mech}.

For typical, relatively short observations of the planet Venus, it can be assumed that the angular velocity of the mirror is constant.
This velocity depends on the spatial positioning of the camera obscura elements 
 -- \rrr{teren}.
Particularly when rays $r_1$ and $r_2$  
are parallel to the equatorial plane, the mirror's angular velocity should be half of the Earth's rotational speed.  
If $r_2$ is parallel to the Earth's rotation axis, the angular velocity should equal the Earth's rotational speed.
Other orientations require intermediate speeds.

%

The described drive mechanism, excluding the motor with gear and controller, is very simple. Additionally, there is an effect of automatic backlash reduction (the elements are tensioned by a spring -- \rrr{mech}) and the advantageous phenomenon of lapping imprecisely machined screw and gear wheel.
%
In simplified form, such a mechanism  could have been built in ancient or medieval times.
The wheel with the screw could be manually rotated at an approximately constant speed. 

We can also imagine a system with feedback in which a person who rotating the wheel receives information about the required speed from an observer seeing the Venus image on the camera obscura screen  (autoguider).  
In another variant of an ``ancient guiding system'', 
a disk with a diaphragm could be moved instead of the mirror. The screw mechanism used for this purpose can be less precise.



\section{Using the camera obscura for photographic observations}
\label{foto}
\nopagebreak
To photograph the planet, we need to replace the screen with either a digital camera sensor or photographic film.
The first challenge is to position the digital or analog camera in such a way that the image of the planet is projected onto the sensor. 

This problem can be solved thanks to the movable mirror system known from single-lens reflex cameras (SLR). The focusing screen and roof pentaprism used in the SRL are replaced by a small telescope (such an instrument is presented in \cite{urania}).  
Looking through this scope, we can find the planet's light in the aperture of the camera obscura disk and take a photograph (exposure settings are given in tables \ref{ttzd_ven} and  \ref{ttzd_plan}). 

The second challenge is the need to use a large image sensor. A solution could be the usage of a large photographic plate or an analog medium-format SLR camera with image size of $6 \times 6$~cm. 

Another idea is to utilize a sensor with a projection system that reduces the image size -- \rrr{proj}(a). 
\begin{figure}[htb]
\begin{center}
\includegraphics[width=0.61\textwidth]{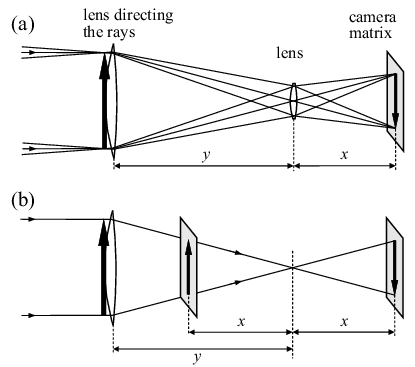}  
\captionss{Projection of the image onto the digital camera sensor.
} 
\label{proj} 
\end{center}
\end{figure}
The projector consists of a lens that scales the image appropriately and a light-directing element.
This part is mounted in the focal plane, where the original image is formed. Its task is to direct the light into the lens aperture (an analogous element is used in the translucent screen -- \rrr{trans}(c)).
Taking advantage of the fact that the rays exiting the camera obscura aperture are nearly parallel, we can omit the lens -- \rrr{proj}(b).
This considerably simplifies the whole device.
The magnification of the optical system can be expressed by the ratio $k = x/y$ (in the case illustrated in \rrr{proj}(a), $k \approx 0.6$).
The value of $x$ can be freely selected because there is no need to adjust the focus (since the rays are almost parallel, the beam of light incident on one point of the image is extremely narrow).
From two possible positions with the same absolute value of magnification, we choose the one in which the sensor is closer to the light-directing element. This makes the entire system more compact. 
The use of the described projector layout not only reduces the image size but also makes the image $1/k^2$ times brighter.

\section{Visual and photographic observations of Venus and other bright planets} 
\label{obser}
\nopagebreak
\subsection{Observations of Venus}
\nopagebreak
A brief description of the conducted visual observations of Venus is presented in \ttt{ttvis_ven}.
%
%
\begin{table*}[htbp]   
\caption{  
\small{ 
Selected visual observations of Venus. In the calculations of $\alpha$ parameter, it was assumed $\lambda=0.55$\eexp{-6}m (observers: 
Krzysztof W\'{o}jcik $^1$ and Adam S\l{}ota $^2$ -- Cracow University of Technology)    
}}   
\setlength{\tabcolsep}{2pt}
\def\arraystretch{1.2}
\begin{center}
\begin{tabular}  
{| @{\tseb}  c  @{\tsea}  | 
   @{\tsec}  c  @{\tsea}  |       
   @{\tseb}  c  @{\tsea}  | 
   @{\tseb}  c  @{\tsea}  | 
   @{\tseb}  c  @{\tsea}  | }
\hline  
\pusline{9.8mm}
\mce{No.} & \mce{Obs. time\\ (UTC), \\ Distace \\ from Earth } & 
\mce{Focal length $f$,\\  Diameter $D$, \\ Parameter $\alpha$ }   & 
\mce{Type of screen, \\ Parameter $\delta_{dir}$     } &  \mce{ Comments} \\[1mm]

\hline    
\pusline{9.6mm}
\mce{1} &     
\mce{2017.02.15 \\ 19:03 \\ 62.0\eeexp{6}km } &  
\mce{ 71 m \\ 15.5 mm \\ 6.2}  &
\mce{Translucent screen \\ (PVC film)  \\  $1.6^\circ$}&
\mces{Directly (without mirror), \\ good observation conditions, \\observation of an elongated \\ Venus shape  $^1$
 }\\[1mm]

\hline    
\pusline{7.8mm}
\mce{2} &     
\mce{2018.12.01 \\ 6:25 \\ 65.5\eeexp{6}km } &  
\mce{ 60 m \\ 13.3 mm \\ 5.4}  &
\mce{Translucent screen \\ (PVC film)  \\  $1.6^\circ$}&
\mces{ Intermediate mirror, \\ observation of an elongated \\ Venus shape $^2$  
 }\\[1mm]

\hline    
\pusline{13.9mm}
\mce{3} &     
\mce{2020.04.22 \\ 20:40 \\ 71.0\eeexp{6}km } &  
\mce{ 69 m \\ 12 mm \\ 3.8}  &
\mce{Reflective screen \\ (aluminum mirror \\ with embossed \\spherical indentations \\ 0.5 mm in diameter) \\  $1.7^\circ$}&
\mces{Directly, very good observation \\conditions, observation of an \\elongated and bulgy \\ Venus shape  $^1$
 }\\[1mm]

\hline    
\pusline{9.6mm}
\mce{4} &     
\mce{2020.05.08 \\ 19:55 \\ 55.0\eeexp{6}km } &  
\mce{ 53 m \\ 12 mm \\ 4.9}  &
\mce{Translucent and \\ reflective screens \\ (like above) \\  $1.6^\circ$  and $1.7^\circ$, resp. }&
\mces{Directly, good observation\\ conditions, observation of an \\ elongated  Venus shape  $^{1, \, 2}$
 }\\[1.25mm]
\hline

\end{tabular}
%
\label{ttvis_ven}
\end{center}
\end{table*}
%
%
In observation no. 3, a tube with a reflective screen was used (see \rrr{arti}). 
To improve the visibility of the image corners, the screen had a concave shape.
In this configuration, an effect similar to that achieved by using a light-directing lens is obtained (the focal length of the concave mirror is approximately 310 mm).

Figure\sss{zd_ven} shows photographs of Venus taken under  various conditions and with different methods.
Brief descriptions are included in \ttt{ttzd_ven}.
Original files and supplementary materials are stored in the repository \cite{rep}.
%
%
\begin{table*}[htbp]   
\caption{  
\small{ 
Photographic observations of Venus, the numbering of the photos corresponds to \rrr{zd_ven} (observers: Krzysztof W\'{o}jcik  $^1$ and Marcin Piekarczyk  $^2$ -- AGH University, Cracow)   
}}   
\setlength{\tabcolsep}{4pt}
\def\arraystretch{1.2}
\begin{center}
\begin{tabular}  
{|@{\tseb}  c  @{\tsea}  | 
  @{\tsec}  c  @{\tsea}  |       
  @{\tseb}  c  @{\tsea}  | 
  @{\tseb}  c  @{\tsea}  | 
  @{\tseb}  c  @{\tsea}  | }  
\hline  
\pusline{9.8mm}
\mce{No.} & \mce{Obs. time\\ (UTC), \\ Distace \\ from Earth } & 
\mce{Focal length $f$,\\  Diameter $D$, \\ Parameter $\alpha$ }   & 
\mce{Exposure time, \\ Sensor sensiv., \\ Projector \\ magnification   } &  \mce{ Comments} \\[1mm]

\hline    
\pusline{7.8mm}
\mce{1} &     
\mce{2017.02.19 \\ 19:27 \\ 62.2\eeexp{6}km } &  
\mce{ 72 m \\ 15.5 mm \\ 6.1}  &
\mce{1/4 s \\ ISO 25000 \\ 1/2.25}&
\mces{Directly (without using an intermediate \\ mirror), exceptionally  good \\ observation conditions $^1$  
 }\\[1mm]

\hline    
\pusline{9.6mm}
\mce{2} &     
\mce{2017.09.01 \\ 2:23 \\ 201\eeexp{6}km } &  
\mce{ 70 m \\ 15 mm \\ 5.8}  &
\mce{1/2 s \\ ISO 25000 \\ 1 (without \\ projector) }&
\mces{ Directly (without mirror), \\ 
the disk with a single aperture \\ was replaced by a 28-hole board,  \\ too high $\alpha$ value $^1$ 
 }\\[1.3mm]

\hline    
\pusline{7.8mm}
\mce{3} &     
\mce{2018.11.29 \\ 5:28 \\ 59.4\eeexp{6}km } &  
\mce{ 104 m \\ 13.3 mm \\ 3.1}  &
\mce{1/13 s \\ ISO 3200 \\ 1/18.9}&
\mces{Intermediate mirror, \\optimal $\alpha$ value $^{1, \, 2}$
 }\\[1mm]

\hline    
\pusline{7.8mm}
\mce{4} &     
\mce{2018.12.09 \\ 3:08 \\ 69,5\eeexp{6}km } &  
\mce{ 54 m \\ 12 mm \\ 4.8}  &
\mce{1/15 s \\ ISO 4500 \\ 1/16}&
\mces{Directly, clearly visible effect \\of light dispersion  in the \\ Earth's atmosphere $^1$
 }\\[1mm]

\hline    
\pusline{7.8mm}
\mce{5} &     
\mce{2018.12.09 \\ 4:11 \\ 69,5\eeexp{6}km } &  
\mce{ 54 m \\ 12 mm \\ 4.8}  &
\mce{1/15 s \\ ISO 4500 \\ 1/16}&
\mces{Intermediate mirror, the observation \\ was made approximately 1 hour later $^1$ 
 }\\[1mm]
\hline

\end{tabular}
%
\label{ttzd_ven}
\end{center}
\end{table*}
%
%
 
%
\begin{figure*}[htb]
\begin{center}
\includegraphics[width=1.0\textwidth]{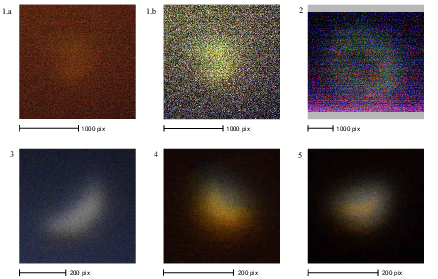}    
\captionss{Photographs of Venus taken with the help of the camera obscura using various techniques, see \ttt{ttzd_ven} (Nikon D7100  DSLR was used, 1 pixel corresponds to 3.9 $\upmu$m). 
Image no. 1.b was created from the original image no. 1.a by increasing the contrast 3.5 times and adjusting the white balance. In image no. 2, the contrast was increased 15 times.} 
\label{zd_ven} 
\end{center}
\end{figure*}
Photographs 1 and 2 (\rrr{zd_ven}) illustrate the difficulty of exposing low-brightness images (image no. 2 was taken without the projector).
To increase the brightness, a relatively large aperture was used, which led to reduced resolution and the formation of a characteristic dark region inside the planet's image (see \roz{resol}).

\subsection{Observations of bright planets }
\nopagebreak
The experience gained while building a camera obscura intended for observing Venus was also applied to observations of other bright planets: Mars, Jupiter, and Saturn.
In this case a key challenge is the significantly lower surface brightness of these planets.  
Their visual observation requires the use of  translucent screens with a very small  beam spread angle ($\delta_{dir} < 2^\circ$). This means that a light-directing element should be used.
A description of the conducted visual observations is provided in \ttt{ttvis_plan}.

Taking a photograph requires precise compensation for the Earth's rotation (e.g., by using an intermediate mirror) and a projection system that increases image brightness.
Alternatively, a large CCD/CMOS sensor with an exposure time of several tens of seconds can be used.
Photographs of Mars, Jupiter, and Saturn are shown in \rrr{zd_plan}. Table \ref{ttzd_plan} describes the conditions and parameters 
under which these photos were taken.
%
%
\begin{table*}[htbp]   
\caption{  
\small{ 
Visual observations of bright planets (observers: 
  Krzysztof W\'{o}jcik $^1$, Adam S\l{}ota $^2$, and Bartosz Zakrzewski $^3$ -- University of the National Education Commission, Cracow, Mt. Suhora Astronomical Observatory)
}}   
\setlength{\tabcolsep}{4pt}
\def\arraystretch{1.2}
\begin{center}
\begin{tabular}  
{ | @{\tseb}  c @{\tsea}  | 
    @{\tsec}  c @{\tsea}  |      
    @{\tseb}  c @{\tsea}  |
    @{\tseb}  c @{\tsea}  | 
    @{\tseb}  c @{\tsea}| }
\hline  
\pusline{9.8mm}
\mce{No.,\\Planet} & \mce{Obs. time\\ (UTC), \\ Distace \\ from Earth } & 
\mce{Focal length $f$,\\  Diameter $D$, \\ Parameter $\alpha$ }   & 
\mce{Type of screen, \\ Parameter $\delta_{dir}$     } &  \mce{ Comments} \\[1mm]

\hline    
\pusline{9.9mm}
\mce{1 \\ Jupiter} &     
\mce{2020.09.08 \\ 20:02 \\ 678\eeexp{6}km } &  
\mce{ 95 m \\ 13.3 mm \\ 3.4}  &
\mce{Translucent screen \\ (PVC film) with \\ light-directing lens \\  $0.6^\circ$}&
\mces{Intermediate mirror, \\ good observation \\conditions  $^1$
 }\\[1mm]

\hline    
\pusline{7.8mm}
\mce{2 \\ Jupiter} &     
\mce{2020.09.08 \\ 20:10 \\ 678\eeexp{6}km } &  
\mce{ 95 m \\ 13.3 mm \\ 3.4}  &
\mce{ Light-directing lens \\ (without PVC film)}&
\mces{Intermediate mirror, \\ good observation \\ conditions  $^{1 \, 3}$
 }\\[1mm]

\hline    
\pusline{9.6mm}
\mce{3 \\ Saturn} &     
\mce{2020.09.08 \\ 20:18 \\ 1.395\eeexp{9}km } &  
\mce{ 95 m \\ 13.3 mm \\ 3.4}  &
\mce{ Light-directing lens }&
\mces{Intermediate mirror, \\ good observation \\ conditions, visible sha-\\pe of Saturn's rings $^{1}$
 }\\[1.2mm]

\hline    
\pusline{7.8mm}
\mce{4 \\ Mars} &     
\mce{2020.09.14 \\ 23:05 \\ 67.3\eeexp{6}km } &  
\mce{ 75 m \\ 16.5 mm \\ 6.6}  &
\mce{ Light-directing lens}&
\mces{Intermediate \\ mirror $^{1 \, 2}$
 }\\[1mm]
\hline

\end{tabular}
%
\label{ttvis_plan}
\end{center}
\end{table*}
%
%
%
\begin{figure*}[htb]
\begin{center}
\includegraphics[width=1.0\textwidth]{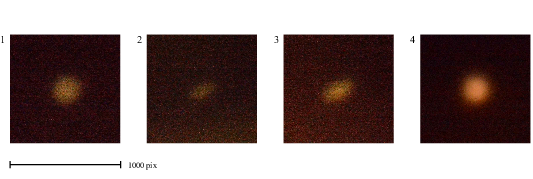}     
\captionss{Photographs of selected bright planets taken using the camera obscura (Nikon D7100), Jupiter (1), Saturn (2,3), and Mars (4).} 
\label{zd_plan} 
\end{center}
\end{figure*}
%
%

%
%
\begin{table*}[htbp]   
\caption{  
\small{ 
Photographic observations of selected bright planets, the numbering of the photos corresponds to \rrr{zd_plan} 
(observers:  
Krzysztof W\'{o}jcik $^1$,
Adam S\l{}ota $^2$, and
Bartosz Zakrzewski $^3$)   
}}   
\setlength{\tabcolsep}{4pt}
\def\arraystretch{1.2}
\begin{center}
\begin{tabular}  
{| @{\tseb}  c @{\tsec} | 
   @{\tsec}  c @{\tsea} |     
   @{\tseb}  c @{\tsea} | 
   @{\tseb}  c @{\tsea} | 
   @{\tseb}  c @{\tsea} | }
\hline  
\pusline{9.8mm}
\mce{No.,\\Planet} & \mce{Obs. time\\ (UTC), \\ Distace \\ from Earth } & 
\mce{Focal length $f$,\\  Diameter $D$, \\ Parameter $\alpha$ }   & 
\mce{Exposure time, \\ Sensor sensiv., \\ Projector \\ magnification   } &  \mce{ Comments} \\[1mm]

\hline    
\pusline{7.8mm}
\mce{1 \\ Jupiter} &     
\mce{2020.09.03 \\ 19:28 \\ 669\eeexp{6}km } &  
\mce{ 95 m \\ 13.3 mm \\ 3.4}  &
\mce{1/8 s \\ ISO 25000 \\ 1/16}&
\mces{Intermediate mirror, \\ good observation \\ conditions  $^{1, \, 2}$
 }\\[1mm]

\hline    
\pusline{7.8mm}
\mce{2 \\ Saturn} &     
\mce{2020.09.03 \\ 20:17 \\ 1.386\eeexp{9}km } &  
\mce{ 95 m \\ 13.3 mm \\ 3.4}  &
\mce{1/2 s \\ ISO 25000 \\ 1/16}&
\mces{Intermediate mirror, \\ good observation\\ conditions  $^{1, \, 2}$
 }\\[1mm]

\hline    
\pusline{7.8mm}
\mce{2 \\ Saturn} &     
\mce{2020.09.08 \\ 20:45 \\ 1.395\eeexp{9}km } &  
\mce{ 100 m \\ 13.3 mm \\ 3.2}  &
\mce{1.3 s \\ ISO 25000 \\ 1/16}&
\mces{Intermediate mirror  $^{1, \, 3}$
 }\\[1mm]

\hline    
\pusline{7.8mm}
\mce{4 \\ Mars} &     
\mce{2020.09.22 \\ 21:21 \\ 64.3\eeexp{6}km } &  
\mce{ 122 m \\ 16.5 mm \\ 4.1}  &
\mce{1/2 s \\ ISO 10000 \\ 1/16}&
\mce{Intermediate mirror, \\ very long focal \\length  $^{1, \, 3}$
 }\\[1mm]
\hline  

\end{tabular}
%
\label{ttzd_plan}
\end{center}
\end{table*}
%
%
%
%
As could have been predicted through straightforward calculations, the visual and photographic observations of Mars and Jupiter depict blurred planetary disks significantly larger than the image of a point source of light. 
However, it is somewhat surprising that the shape of Saturn's rings is relatively easy to discern during visual observation.

\section{A brief overview of the results}  
 \label{discussion}
\nopagebreak
%
The results of the observations reported indicate that it is feasible to observe the crescent shape of Venus with a camera obscura.
This applies to both visual and photographic observations.
A crucial factor enabling visual observation was the use of a directional screen, particularly a translucent one (\roz{bright}).
A simple guiding system is also very helpful (\roz{comp}).


The structure  of the camera obscura is scalable.
Theoretically, by increasing its size, any desired resolution can be achieved.
The camera in which the projection system is used shows, however, something more. 
In this layout, two functions of the camera working as a telescope are separated, namely: 
\begin{itemize}
\item detecting the direction from which the light of the examined object comes,
\item focusing its energy. 
\end{itemize}

The first function determines the resolving power. 
In a camera obscura, this depends on its geometry and diffraction phenomena (\ref{rt}). 
Apart from these limitations, there are no factors that reduce the camera's resolution, such as imperfections in mirror or lens surfaces, temperature effects, etc.

The second function (energy focusing) is performed by the projection system. This part is also scalable.
Moreover, as a result of separating the two functions desctibed, the projection system does not need to direct the light with high precision.

In a large-scale ``cosmic'' camera obscura (e.g., $D = 140$~m, $f = 10$\eeexp{6}km), 
the image size could be on the order of kilometers.
One can imagine constructing a projection device of this size either on Earth or in outer space.
Such a system could, for example, consist of many mirrors located in the screen plane,
with each mirror corresponding to a pixel or group of pixels.
Of course, in such a telescope, the issue of  pointing it at specific objects remains unresolved.
Its operation would likely be limited to slowly scanning regions of the sky at which it is passively directed,
as a result of the movement of its components in space.

\section{Summary}  
 \label{concl}
\nopagebreak
%
%

The main issue addressed in this article is the observation of the phases of Venus using a camera obscura.
The results of the observations presented in \roz{obser} demonstrate that such observations can be successfully carried out.
The paper also provides a detailed description of the cameras used in these experiments.
We can therefore conclude that the main goals of the work articulated in \roz{intro} have been achieved.     

To reach these research goals, several key methods and solutions were employed, namely:
\begin{enumerate}
\item 
The use of a large-scale camera obscura (with a focal length exceeding 50 m).
\item 
The use of a directional screen (using the translucent screens with a small angle of beam spread is particularly effective).
\item 
The implementation of a simplified, even manually controlled, guiding system
(the proposed screw-based design is simple enough to be built using relatively primitive methods).
\item
The separation of the two functions of the camera obscura that works as a telescope:  detecting the direction of incoming light and focusing its energy.
The components that perform these functions (i.e., the aperture disk and projection system) are fully scalable.
\end{enumerate}

Observations of Venus using a camera obscura can be applied in the education and popularization of astronomy.
They can contribute greatly to the understanding of the fundamental laws of optics and issues related to planetary motion. 
Conducting such observations fosters patience, develops practical skills in using basic instruments, and inspires creativity in designing new versions of observational tools.

Let us return to the question of whether it was possible to observe the crescent of Venus using a camera obscura in ancient or medieval times.
Based on the observation methods described in this article, it can be concluded that such observations could indeed have been conducted.
A series of observations of Venus's phases would reveal that the planet is a sphere illuminated by sunlight.
As mentioned earlier, a correct interpretation of this fact could have led to the formulation and validation of the heliocentric model of the universe.


The question of whether our ancestors actually employed these methods is of great significance for the history of astronomy, the history of science,
and human thought in general.
Thus, there is a need to conduct additional archeological and historical studies.  
The author hopes that the instruments and observation techniques described in this paper may narrow down the scope of this research.

For instance, it is easy to recognize that large, stair-shaped buildings could be used to construct camera obscuras.
Their specific architectural design facilitates the mounting and adjustment of the aperture disc and observation tube.
One can easily point to examples of such buildings and even pairs of them, 
positioned in alignment with the cardinal directions, e.g., the Pyramid of Kukulkan and the entire complex of buildings in Chich\'{e}n Itz\'{a} (Mexico), the Pyramid of the Sun in Teotihuac\'{a}n (Mexico), or the Egyptian pyramids.




\vspace{4pt} 




\end{document}